\documentclass[aoas,nameyear,dvips]{arximspdf}
\usepackage{dcolumn}
\usepackage{graphicx}

\doi{10.1214/09-AOAS276}
\volume{4}
\issue{1}
\pubyear{2010}
\firstpage{422}
\lastpage{438}

\makeatletter
\DeclareMathAlphabet\mathcaligr{OMS}{cmsy}{m}{n}
\newcolumntype{d}[1]{D{.}{.}{#1}}
\renewcommand{\citep}[1]{[\citet{#1}]}
\renewcommand{\cite}[1]{\citet{#1}}
\def\E{\mathrm{E}}
\def\P{P}
\def\Var{\operatorname{Var}}
\newtheorem{lem}{Lemma}
\makeatother

\begin{document}
\begin{frontmatter}

\title{An empirical Bayes mixture method for effect size and false discovery
rate estimation}
\runtitle{An empirical Bayes mixture method for effect size and FDRS}

\begin{aug}
\author{\fnms{Omkar} \snm{Muralidharan}\thanksref{a1}\ead[label=e1]{omkar@stanford.edu}\corref{}}
\runauthor{O. Muralidharan}
\affiliation{Stanford University}
\address{Department of Statistics\\
Stanford University\\
390 Serra Mall\\
Stanford, California 94305\\
USA\\
\printead{e1}} 

\thankstext{a1}{Supported by an NSF VIGRE Fellowship.}
\end{aug}

\received{\smonth{5} \syear{2009}}
\revised{\smonth{7} \syear{2009}}

%
\begin{abstract}
Many statistical problems involve data from thousands of parallel
cases. Each case has some associated effect size, and most cases will
have no effect. It is often important to estimate the effect size
and the local or tail-area false discovery rate for each case. Most
current methods do this separately, and most are designed for normal
data. This paper uses an empirical Bayes mixture model approach to
estimate both quantities together for exponential family data. The
proposed method yields simple, interpretable models that can still
be used nonparametrically. It can also estimate an empirical null
and incorporate it fully into the model. The method outperforms existing
effect size and false discovery rate estimation procedures in normal
data simulations; it nearly acheives the Bayes error for effect size
estimation. The method is implemented in an R package (mixfdr), freely
available from CRAN.
\end{abstract}

%

\end{frontmatter}

Suppose we have $N$ parallel cases, each with some effect size $\delta_{i}$.
We observe a measurement $z_{i}\sim f_{\delta_{i}}$ independently
for each case. We want to estimate how big each effect is and narrow
in on the few cases of interest. To do this, we must estimate $\delta_{i}$
and either the local false discovery rate, $\mathit{fdr}(z)=\P(\delta_{i}=0|z_{i})$,
or the tail-area false discovery rate, $\mathit{FDR}(z)=P(\delta
_{i}=0||z_{i}|\geq z)$.
This problem comes up in many different areas: microarrays motivate
this paper, but the question also arises in data mining, model selection
and image processing [\cite{Abramovich2006}, \cite{Abramovich2007}, \cite{Johnstone2004}].

We present a mixture model empirical Bayes method to solve this problem
in Section~\ref{sec:The-Model}. A simple hierarchical model lets
us estimate effect sizes and false discovery rates in a flexible,
conceptually neat way. The approach works for general exponential
families $f_{\delta}$, and can estimate an empirical null. We illustrate
the method for binomial data in Section~\ref{sec:Binomial-Data-Example}.
Simulation results in Section~\ref{sec:Normal-Data-Simulations} show
that the method performs well on normal data: it estimates $\delta$
nearly as well as the Bayes rule, and is a better $\mathit{fdr}$ estimator
than existing methods.

\section{Model}\label{sec:The-Model}

Our model is a specialization of the Brown--Stein model used by \cite
{Efron2008a}.
This model supposes $(\delta_{i},z_{i})$ are independently generated
by the following hierarchical sampling scheme:
\begin{eqnarray*}
\delta
& \sim& g(\delta),\\
z|\delta& \sim& f_{\delta}(z),
\end{eqnarray*}
where $f_{\delta}(z)$ is an exponential family with natural parameter
$\delta$. Given the prior $g$, we can calculate $\mathit{fdr}(z)$, $\mathit{FDR}(z)$
and the Bayes estimator of $\delta$, $\E(\delta|z)$. However, we
usually do not want to specify $g$ in advance. Instead, we can take
an empirical Bayes approach: use the data to estimate $g$, and use
this estimated prior to get effect size and false discovery rate estimates.

\subsection*{Mixture prior}
Modeling $g$ as a mixture gives us the flexibility of a nonparametric
model for $g$ with the convenience and stability of a parametric
one. We model $g$ as a mixture of $J$ priors $g_{j}$:
%
\begin{equation}\label{eq:mixmodel}
g(\delta)=\sum_{j=0}^{J-1}\pi_{j}g_{j}(\delta).
\end{equation}
The priors $g_{j}$ are taken from some parametric family of priors
for $\delta$, and each has a hyperparameter vector $\theta_{j}$.
We usually think that the marginal distribution of $z$, $f(z)$,
has a known null component $f_{0}$, corresponding to the many cases
with $\delta=0$. To model this, we think of the $0$th mixture component
as null, and fix $\theta_{0}$ so that $g_{0}$ is a point mass at
$0$. The other parameters $\theta_{j}$ and the mixture proportions
$\pi_{j}$ are unknown, and must be estimated. We fit them using marginal
maximum likelihood via the EM algorithm. We can also incorporate case-specific
nuisance parameters into the model as long as they can be estimated.
Details for these issues are given in the supplementary information
\citep{Muralidharan2009}.

We can choose any family of priors as long as we can calculate the
posteriors, and the family is rich enough to model $g$ nonparametrically
given enough components. With such a family, we can go from a strongly
parametric model to a nearly nonparametric model by increasing $J$.
It is often very convenient to work with conjugate priors for
$f_{\delta}$,
since the posterior distributions are easy to calculate.

The mixture model gives the posterior distribution of $\delta|z$
a simple form, making it easy to calculate $\mathit{fdr}(z)$ and $\E(\delta|z)$.
Let $f^{(j)}=\int f_{\delta}g_{j}(\delta)\,d\delta$ be the $j$th group
marginal, so the marginal distribution of $z$ is $f(z)=\sum\pi_{j}f^{(j)}(z)$,
and let $F^{(j)}$ and $F$ be corresponding cdfs (the superscripts
are to avoid confusion with $f_{\delta}$). Let $p_{j}(z)=\frac{\pi
_{j}f^{(j)}(z)}{f(z)}$
be the posterior probability that $z$ came from group $j$, and
$g_{j}(\delta|z)$
be the posterior for the $j$th group (that is, the posterior corresponding
to prior $g_{j}$). Then under model~(\ref{eq:mixmodel}), the posterior
distribution is a mixture:
%
\begin{equation}\label{eq:posterior}
\delta|z\sim\sum_{j=0}^{J-1}p_{j}(z)g_{j}(\delta|z).
\end{equation}
In particular, this gives us our estimates:
\begin{eqnarray*}
\mathit{fdr}(z)
& =&
p_{0}(z),\\
\mathit{FDR}(z) & =&\frac{\pi_{0} (1-F^{(0)}(z)+F^{(0)}(-z)
)}{1-F(z)+F(-z)},\\
\E(\delta|z) & =&\sum_{j=0}^{J-1}p_{j}(z)\E_{j}(\delta|z),
\end{eqnarray*}
where $\E_{j}$ denotes the expectation under $g_{j}(\delta|z)$.
Other quantities, like the posterior variance $\Var(\delta|z)$, can
be calculated easily using equation~\ref{eq:posterior}. These formulas
are derived in the supplementary information \citep{Muralidharan2009}.

\subsection*{Empirical nulls}

This model can accommodate empirical nulls by penalizing the mixture
proportions and allowing the null component $g_{0}$ to vary. Sometimes,
because of correlation or other issues, it is no longer true that
most $z\sim f_{0}$ \citep{Efron2008}. This makes the theoretical
null inappropriate; instead, Efron suggests fitting an empirical null
so that most $z$ have the empirical null distribution. In the mixture
model, using an empirical null corresponds to $g_{0}$ not being a
point mass at $0$ and $\pi_{0}$ being larger than the other $\pi$'s.
We can therefore fit an empirical null by letting $g_{0}$ vary and
putting a penalty on the proportions $\pi$. The most convenient and
interpretable penalty corresponds to a $\operatorname{Dirichlet}(\beta)$ prior on
$\pi$. These modifications are easy to incorporate into the fitting
process (details are in the supplementary information \citep{Muralidharan2009}).
Penalizing $\pi$ is useful even for the theoretical null---it stabilizes
the parameter estimates by mitigating the effect of the likelihood's
multiple local maxima.

\subsection*{Tuning parameters and how to choose them}

This method has two tuning parameters---the penalization parameter
$\beta$ and the number of mixture components~$J$. Perhaps somewhat
counterintuitively, $J$ is less important and easier to choose. This
is because for typical datasets, it has little effect on the fitted
density~$f$, and $\E(\delta|z)$ is a function of $f$ (as Lemma
\ref{lem:fdr} will show). If we treat nearly null components as null
(see the next subsection), $\mathit{fdr}$ and $\mathit{FDR}$ estimates are insensitive
to $J$ as well. The literature on mixture models has many methods
to choose~$J$ \citep{McLachlan2000}; one easy method is to use the
Bayes Information Criterion. For most purposes, however, we can just
fix $J$. Taking $J=3$ works particularly well. This choice gives
a group each to null, positive effect and negative effect cases.

The penalization $\beta$ can be more important. It is usually best to
choose $\beta=(P,0,0,\ldots,0)$. With this choice, the exact value
of $P$ is not important for effect size estimation and $\mathit{fdr}/\mathit{FDR}$
estimation with the theoretical null. With empirical nulls, however,
$P$ can be more important. A larger $P$ forces a bigger null group,
and so increases estimates of the null variance. This can have a big
effect on $\mathit{fdr}$ estmates.

We can choose $P$ with a simple parametric bootstrap calibration
scheme. First list some candidate penalizations $P_{1},\ldots,P_{K}$
(usually $20$ penalizations evenly spaced between $100$ and $\frac{N}{2}$).
Then, fit a preliminary model $m$ to the data using some reasonable
default penalization ($P=\frac{1}{5}N$ is a good choice). Next, create
perturbed models $m_{1},\ldots,m_{L}$ by changing the null parameters
slightly, and possibly changing the alternatives. We will choose $P$
to be the $P_{k}$ that performs best over the perturbed models. To
assess performance, generate $B$ random data sets of size~$N$ from
each $m_{l}$. Fit $k$ mixture models to each bootstrap data set,
one for each penalization $P_{k}$, and see how close each of the
fitted models is to the true model for that data set (which will be
one of the $m_{l}$'s). The best $P$ is the one that performs best
over all the bootstrap data sets.

It is worth emphasizing, however, that the mixture model is relatively
insensitive to parameter choice. Both $J$ and $P$ have little effect
on the fitted density, and so do not affect effect size and theoretical
null $\mathit{fdr}/\mathit{FDR}$ estimates too much. This is seen in the simulations
of Section~\ref{sec:Normal-Data-Simulations}, where the mixture model
nearly acheives the Bayes effect size estimation error for many different
combinations of $J$ and $P$.

\subsection*{Choosing a null hypothesis}

The mixture model also raises a new question: how should we treat
nearly null mixture components? Fitting often gives mixture components
that are nearly, but not quite, null. For example, $g_{1}$ might
not be a point mass at $0$, but still give $\delta$ close to $0$
with high probability. We need to decide whether to include these
components in the null when estimating $\mathit{fdr}$'s and $\mathit{FDR}$'s. \cite{Efron2004a}
argues that the answer depends on whether the nearly null components
are still interesting in the presence of strongly null components.
The nearly null components, however, are usually highly sensitive
to tuning parameters---different parameters can change the nearly
null components dramatically with little effect on the overall density
$f$. It is thus usually best to include the nearly null components
in the null. If the components are insensitive to parameter choice,
though, Efron's answer is correct, and the question becomes a scientific
one.

\subsection*{Identifiability concerns}

One problem with this method is that mixture models can be nearly
unidentifiable. We can have very different models for $g$ that give
nearly the same marginal $f$. We cannot choose between such models
based on the data, so estimates of $g$ cannot always be taken seriously.
The following result, however, shows that the mean and variance of
the posterior distribution $g(\delta|z)$ are simple functions of
$f$, and thus can be taken seriously. The result is a generalization
of Efron's calculations in \cite{Efron2008a} to exponential families,
though the formula goes back to \cite{Robbins1954}. It applies for
the Brown--Stein model in general, not just to the mixture model.
\begin{lem}\label{lem:fdr}
Assume we are in the Brown--Stein model for exponential
families and $z$ is continuous. Then the mean and variance of the
posterior distribution $g(\delta|z)$ are given by
\begin{eqnarray*}
\E(\delta|z) & =&-\frac{d}{dz} \biggl(\log\frac
{f_{0}(z)}{f(z)} \biggr),\\
\Var(\delta|z) & =&-\frac{d^{2}}{dz^{2}} \biggl(\log\frac
{f_{0}(z)}{f(z)} \biggr).
\end{eqnarray*}

If we use the theoretical null and $\pi_{0}$ is known, then
$\mathit{fdr}(z)=\frac{\pi_{0}f_{0}(z)}{f(z)}$
and $\mathit{FDR}(z)=\frac{\pi_{0} (1-F_{0}(z)+F_{0}(-z) )}{1-F(z)+F(-z)}$
are also functions of $f(z)$.
\end{lem}
\begin{pf}
The proof follows \citep{Efron2008a} closely. Recall that in the
Brown--Stein model we assume only that $\delta$ has prior $g(\delta)$,
and $z|\delta\sim f_{\delta}$. The posterior of $\delta|z$ is
\begin{eqnarray*}
g_{\delta|z}(\delta) & = & \frac{f_{\delta}(z)g(\delta)}{f(z)}\\
& = & \exp \biggl(z\delta-\log\frac{f(z)}{f_{0}(z)} \biggr)e^{-\psi
(\delta)}g(\delta).
\end{eqnarray*}
Thus, $\delta|z$ is distributed according to an exponential family
with natural parameter~$z$ and cumulant generating function $-\log
\frac{f_{0}(z)}{f(z)}$.
The cumulants of $\delta|z$ are immediately obtained by differentiating
this function. Note that this proof goes through for multiparameter
exponential families as well.
\end{pf}

Lemma~\ref{lem:fdr} connects effect size and $\mathit{fdr}$ estimation in
exponential families, and is thus useful beyond the mixture model---any density (or equivalently, $\mathit{fdr}$) estimation method gives us
effect size estimates. Such an approach is even useful for discrete
families, where the lemma does not apply. The proof shows that
$g_{\delta|z}$
is well defined for $z$ in some convex set that includes the sample
space of $z$. The problem in the discrete case is that we only know
the value of the cumulant generating function in the sample space,
and this is not enough to differentiate. We can, however, estimate
the cgf by interpolating the known or estimated values. Differentiating
this gives us estimates of $\E(\delta|z)$ and $\Var(\delta|z)$ corresponding
to priors whose posterior cgfs are not too wild. This method performs
well on simulated binomial and Poisson data despite its somewhat shaky
theoretical foundations.

\subsection*{Connections to existing methods}

This model differs from most $\mathit{fdr}$ and effect size estimation methods
in three important ways. First, it estimates $\mathit{fdr}$'s and effect sizes
together, not separately. Second, it incorporates its empirical null
estimate into its overall density estimate. Finally, it works in general
exponential families, not just for normal data or $p$-values.

That said, this mixture model is closely connected to many existing
$\mathit{fdr}$ and effect size estimation procedures. $\mathit{fdr}$ estimation under
the theoretical null reduces to estimating $\pi_{0}$ [see, for example,
\cite{Storey2002}, \cite{Cai2007}, \cite{Jin2007}, \cite{Meinshausen2006}] and $f$ [examples
include \cite{Efron2008}, \cite{Strimmer2008}] since $\mathit{fdr}=\frac{\pi_{0}f_{0}}{f}$
[\cite{Efron2001}, \cite{Storey2002}]. In this context, the proposed method
corresponds to using a mixture model density estimation method. This
approach has been successfully used for normal data [\cite{Pan2003}, \cite{McLachlan2000}],
$p$-values \citep{Allison2002} and Gamma data \citep{Newton2004}.
In particular, our treatment of empirical nulls is similar to that
of \cite{McLachlan2000}. The proposed method goes further than these
methods by incorporating an empirical null estimate into the density
estimate and using the mixture model to estimate effect sizes.

The proposed method is also similar to many effect size estimation
procedures. Many effect size estimation methods use a two group mixture
model for $g$ and estimate $\delta$ with the posterior mean, median
or mode. The model can either be specified in advance or estimated
empirically---both approaches can yield theoretically attractive estimators
[\cite{Johnstone2004}, \cite{Pensky2006}, \cite{Abramovich2007}]. Our mixture model
can be viewed as a particular instance of this general recipe for
effect size estimation, adapted to estimate $\mathit{fdr}$'s as well. The
model is also closely related to another family of procedures that
use density estimates and a normal data version of Lemma~\ref{lem:fdr}
to estimate effect sizes [\cite{Efron2008a}, \cite{Brown2008}]. For continuous
$z$, the proposed method corresponds to using a particular mixture
density estimator and the more general Lemma~\ref{lem:fdr} to transform
the density estimate to an effect size estimate.

\section{Binomial data example}\label{sec:Binomial-Data-Example}

To illustrate the mixture model, we use it to predict Major League
Baseball batting averages. The data consist of batting records for
Major League Baseball players in the 2005 season. We assume that each
player has a true batting average $\delta_{i}$, and that his hit
total $H_{i}$ is $\operatorname{Binomial}(N_{i},\delta_{i})$, where~$N_{i}$ is
the number of at bats. The goal is to estimate each players' batting
average $\delta_{i}$ based on the first half of the season. We restrict
our attention to players with at least $11$ at bats in this period
(567 players).

\subsection*{Brown's analysis}

\cite{Brown2008} analyzes the data using a normalizing and variance
stabilizing transformation. He transforms the data $(H,N)$ to
\[
X_{i}=\arcsin\sqrt{\frac{H_{i}+1/4}{N_{i}+1/2}},
\]
and the transformed data are approximately normal
\begin{eqnarray*}
X_{i} & {\dot{\sim}}&\mathcaligr{N}\biggl(\mu_{i},\frac{1}{4N_{i}}\biggr),\\
\mu_{i} & =& \arcsin\sqrt{\delta_{i}}.
\end{eqnarray*}
He estimates $\mu_{i}$ using the following methods:
\begin{itemize}
\item The naive estimator, $\hat{\mu}_{i}=X_{i}$.
\item The overall mean, $\hat{\mu}_{i}=\bar{X}$.
\item A parametric empirical Bayes method that models $\mu_{i}\sim
\mathcaligr{N}(\mu,\tau^{2})$.
The prior parameters $\mu$ and $\tau$ are fit either by method of
moments or maximum likelihood.
\item A nonparametric empirical Bayes method. First, Brown estimates the
marginal density of each $X_{i}$ with a kernel density estimator
(tweaked because of the unequal variances). Then he uses a normal
version of Lemma~\ref{lem:fdr} from \cite{Brown1971} to estimate
$\mu$.
\item The positive part James--Stein estimator.
\item A Bayesian estimator that models $\mu_{i}\sim\mathcaligr{N}(\mu
,\tau^{2})$,
$\mu\sim \operatorname{Unif}(\mathbb{R})$, $\tau^{2}\sim \operatorname{Unif}(0,\break\infty)$.
\end{itemize}
Finally, Brown estimates the estimation error of these methods using
their prediction error on the second half of the season. Let $(\tilde
{H}_{i},\tilde{N}_{i})$
be the data for the second half of the season. Brown's error criterion
is
%
\begin{equation}\label{eq:TSE}
\mathit{TSE}=\sum (\hat{\mu}_{i}-\tilde{X}_{i} )^{2}-\frac
{1}{4\tilde{N}_{i}}.
\end{equation}
By construction, $\E(\mathit{TSE})=\sum(\hat{\mu}_{i}-\mu_{i})^{2}$. The methods
are assessed over all players who had at least $11$ at bats in each
half of the data (499 players).

\subsection*{Mixture model}

We can analyze the data on the original scale using a binomial mixture
model. We model the data using the Brown--Stein model [$\delta_{i}\sim
g(\delta)$,
$H_{i}|\delta_{i}\sim \operatorname{Binomial}(N_{i},\delta_{i})$], and model $g$
as a mixture of Beta distributions
\[
g(\delta)=\sum_{j=0}^{J}\pi_{j}\operatorname{Be}(\delta;\alpha_{j},\beta_{j}).
\]
This model makes the marginal distribution of $H_{i}$ a mixture of
Beta-binomial distributions, $f(H_{i};N_{i})=\sum\pi_{j}f^{(j)}(H_{i};N_{i})$.
The conjugate property of the Beta prior makes the posterior distributions
simple:
\[
g(\delta_{i}|H_{i})=\sum_{j=0}^{J}p_{j}(H_{i})\operatorname{Be} (\delta
;\alpha_{j}+H_{i},\beta_{j}+N_{i} ),
\]
where $p_{j}(H_{i})=\frac{\pi_{j}f^{(j)}(H_{i};N_{i})}{f(H_{i};N_{i})}$.
The parameters $\pi$, $\alpha$ and $\beta$ are fitted by marginal
maximum likelihood via the EM algorithm (details are in the supplementary
information \citep{Muralidharan2009}). For easy comparison with Brown's
results, we estimate $\mu_{i}$ by its posterior mean $\E(\arcsin
\sqrt{\delta}|z)$.

\subsection*{Results}

Table~\ref{tab:baseres} compares the mixture model to Brown's methods
--- the mixture model is a good performer, but not the best. It performs
about $15\%$ worse than the nonparametric empirical Bayes and James--Stein
estimators. Brown observes that the number of at bats is correlated
with the batting averages---better batters bat more. This violates
all methods' assumptions, but has a particularly strong effect on
the more parametric methods. Splitting the players into pitchers (81
training, 64 test) and nonpitchers (486 training, 435 test) reduces
this effect.

%
\begin{table}
\caption{Estimated estimation accuracy [equation (\protect\ref{eq:TSE})]
for the methods. The naive estimator is normalized to have error $1$.
Values for all methods except the binomial mixture model are from
Brown (\protect\citeyear{Brown2008}). The first column gives the errors on the data
as a whole (single model), and the next two give errors for pitchers
and nonpitchers considered separately. Standard errors range from
$0.05$ to $0.2$ on nonpitchers, are higher for pitchers, and are
in between for the overall data [Brown (\protect\citeyear{Brown2008})]}
\label{tab:baseres}
\begin{tabular*}{\textwidth}{@{\extracolsep{\fill}}ld{1.3}d{1.3}d{1.3}@{}}
\hline
& \multicolumn{1}{c}{\textbf{Overall}} & \multicolumn{1}{c}{\textbf{Pitchers}}
& \multicolumn{1}{c@{}}{\textbf{Nonpitchers}}\\
\hline
\textit{Number of training players} & \multicolumn{1}{c}{\textit{567}}
& \multicolumn{1}{c}{\textit{81}} & \multicolumn{1}{c@{}}{\textit{486}}\\
\textit{Number of test players} & \multicolumn{1}{c}{\textit{499}} & \multicolumn{1}{c}{\textit{64}}
 & \multicolumn{1}{c@{}}{\textit{435}}\\
Naive & 1 & 1 & 1\\
Group mean & 0.852 & 0.127 & 0.378\\
Parametric empirical Bayes (Moments) & 0.593 & 0.129 &
0.387\\
Parametric empirical Bayes (ML) & 0.902 & 0.117 & 0.398\\
Nonparametric empirical Bayes & 0.508 & 0.212 & 0.372\\
Bayesian estimator & 0.884 & 0.128 & 0.391\\
James--Stein & 0.525 & 0.164 & 0.359\\[6pt]
\textbf{Binomial mixture model} & \multicolumn{1}{c}{\textbf{0.588}} &\multicolumn{1}{c}{ \textbf{0.156}} &
\multicolumn{1}{c@{}}{\textbf{0.314}}\\
\hline
\end{tabular*}
\end{table}

The results, also in Table~\ref{tab:baseres}, show that splitting
makes the mixture model the best performer for nonpitchers and an
average performer for pitchers. Splitting also reduces the differences
between the methods. Both the nonparametric empirical Bayes estimator
and the binomial mixture model do relatively better on nonpitchers
than on pitchers. This is probably because the smaller number of pitchers
makes it difficult to estimate the marginal density. Simple simulations
show that the binomial mixture model is probably truly better than
the other methods for nonpitchers, but no firm conclusions can be
drawn about the methods' relative performance on pitchers or the combined
data.

The binomial mixture model has advantages beyond possible performance
gains. It removes the need for a normalizing and variance stabilizing
transformation by working with the original data. It can estimate
any function $h(\delta)$, since $\E(h(\delta)|z)$ can be calculated
numerically. Finally, the mixture prior can be informative. For example,
the estimated prior for nonpitchers was a single $\operatorname{Beta}(302,884)$
distribution, while the estimated pitchers' prior was a mixture of
$\operatorname{Beta}(90,983)$ and $\operatorname{Beta}(219,928)$ distributions. These prior estimates
were stable under different choices of $J$ and starting points for
the EM algorithm. This could indicate that nonpitchers are about
the same across the league, but pitchers come in two different types.

\section{Normal data simulations}\label{sec:Normal-Data-Simulations}

In this section we shall see that the mixture model performs very
well in the important normal case. The mixture model is particularly
simple for normal data. We use the Brown--Stein model [$\delta\sim
g(\delta)$,
$z|\delta\sim\mathcaligr{N}(\delta,1)$] and model the prior $g$ as
a normal mixture:
\[
g(\delta)=\sum_{j=0}^{J-1}\pi_{j}\varphi(\delta;\mu_{j},\sigma_{j}^{2}),
\]
where $\varphi(x;\mu,\sigma^{2})$ is the $\mathcaligr{N}(\mu,\sigma^{2})$
density function. This model makes the marginal $f$ a normal mixture,
$f(z)=\sum\pi_{j}\varphi(z;\mu_{j},\sigma_{j}^{2}+1)$. Fixing $\mu_{0}=0$,\vspace*{1pt}
$\sigma_{0}=0$ corresponds to using a theoretical null, and letting
them vary corresponds to using an empirical null. Normality makes
the posterior $g(\delta|z)$ simple. It is easy to check that
\begin{eqnarray*}
g(\delta|z) & =&\sum_{j=0}^{J}p_{j}(z)\varphi \biggl(\delta;\frac
{1}{\sigma_{j}^{2}+1}\mu_{j}+\frac{\sigma_{j}^{2}}{\sigma
_{j}^{2}+1}z,\frac{\sigma_{j}^{2}}{\sigma_{j}^{2}+1} \biggr),\\
\mathit{fdr}(z) & =&p_{0}(z),\\
\E(\delta|z) & =&\sum p_{j}(z) \biggl(\frac{1}{\sigma_{j}^{2}+1}\mu
_{j}+\frac{\sigma_{j}^{2}}{\sigma_{j}^{2}+1}z \biggr),
\end{eqnarray*}
where $p_{j}(z)=\frac{\pi_{j}\varphi(z;\mu_{j},\sigma_{j}^{2}+1)}{f}$.
The parameters $\pi$, $\mu$ and $\sigma$ are estimated by marginal
maximum likelihood via the EM algorithm. We used a
$\operatorname{Dirichlet}(P,0,\break\ldots,0)$
penalty on $\pi$ to stabilize the model. The normal mixture model
approach is implemented in an R package ``mixfdr,'' available
from CRAN and the author's website.

\subsection*{Effect size estimation}

We can investigate the effect size estimation performance of the normal
mixture model with simulation closely based on one done by \cite
{Johnstone2004}.
We generate $z_{i}\sim\mathcaligr{N}(\delta_{i},1)$, for $i=1,\ldots,N=1000$.
The goal is to estimate $\delta_{i}$ based on $z$ and minimize the
squared error $\sum(\delta_{i}-\hat{\delta}_{i})^{2}$. $K$ of the
$\delta_{i}$ were nonzero. In the one-sided case, the nonzero $\delta_{i}$
were i.i.d. $\operatorname{Unif}(\mu-\frac{1}{2},\mu+\frac{1}{2})$; in the two-sided
case, two-thirds of the $\delta_{i}$ were $\operatorname{Unif}(\mu-\frac{1}{2},\mu
+\frac{1}{2})$
and one-third were $\operatorname{Unif}(-\mu-\frac{1}{2},-\mu+\frac{1}{2})$. Different
values of $K$ and $\mu$ were used to simulate different combinations
of sparsity and effect strengths. We will compare the mixture model
to the following effect size estimation methods:
\begin{itemize}\label{ite:A-spline-density}
\item A spline density method used by \cite{Efron2009}.
\item EBayesThresh, an empirical Bayes approach taken by \cite{Johnstone2004}.
\item SUREShrink, a method based on minimizing Stein's Unbiased Risk Estimate
for thresholding \citep{Donoho1995}.
\item FDR-based thresholding \citep{Abramovich2006}, at threshold $q=0.1$.
\item Soft and hard thresholding using the ``universal threshold''
$\sqrt{2\log N}\approx3.7$
from \cite{Donoho1994}.
\end{itemize}
All methods use the known variance of $z$, and when applicable, assume
a theoretical $\mathcaligr{N}(0,1)$ null. All methods' tuning parameters
were hand-picked for good performance over the simulation scenarios,
but none were rigorously optimized (including the mixture model, which
used $J=10$ and $P=50$). The whole simulation was repeated $100$
times, and the same random noise was used for each scenario and each
method. Code for the simulation, a slightly modified version of the
code used by \cite{Johnstone2004}, is available in the Supplementary
Material online.

%
\begin{figure}[b]

\includegraphics{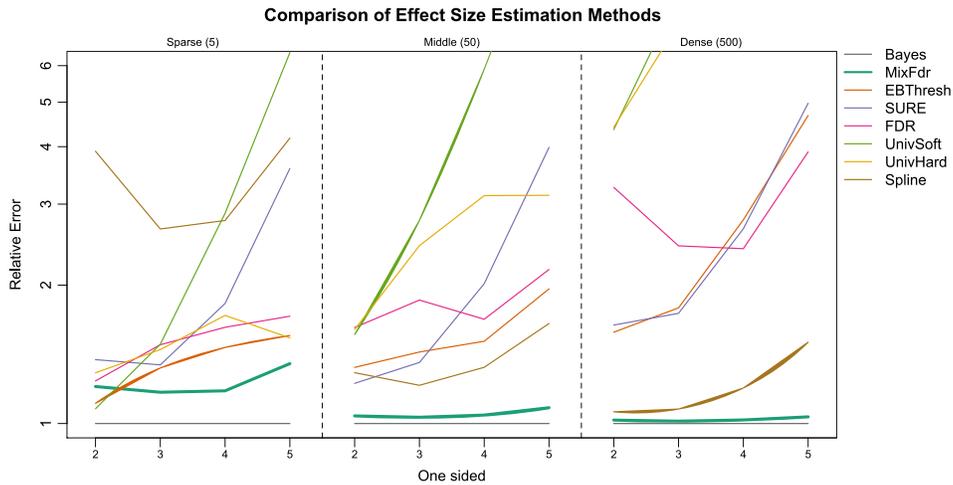}

\caption{Simulation results for the one-sided scenario.
Each panel corresponds to one value of $K$ ($5$, $50$ or $500$).
Within each panel, $\mu$ increases from $2$ to $5$. The $y$-axis
plots the squared error [$\sum(\delta_{i}-\hat{\delta}_{i})^{2}$],
averaged over $100$ replications. Errors are normalized so that the
Bayes estimator for each choice of $K$ and $\mu$ has error $1$.
Estimation methods are listed in the text. In the dense case, the
universal soft and hard thresholding methods are hidden because their
relative errors range from $4$ to $40$.}\label{fig:one-sided}
\end{figure}

\begin{figure}

\includegraphics{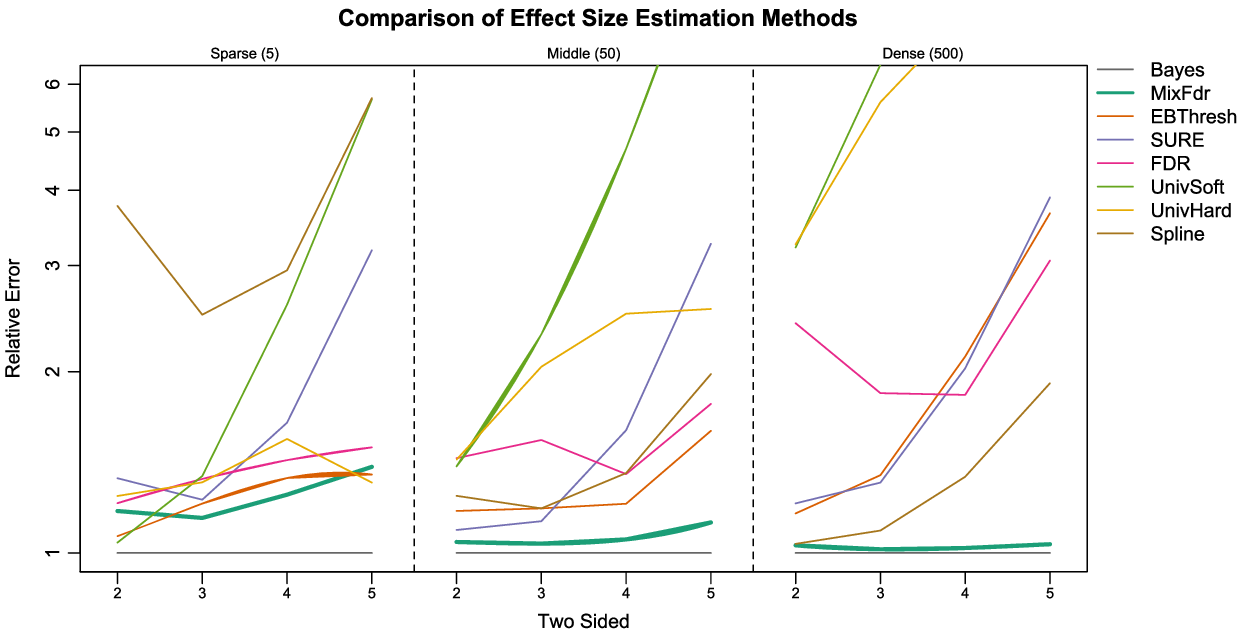}

\caption{Simulation results for the two-sided scenario.
Each panel corresponds to one value of $K$ ($5$, $50$ or $500$).
Within each panel, $\mu$ increases from $2$ to $5$. The $y$-axis
plots the squared error [$\sum(\delta_{i}-\hat{\delta}_{i})^{2}$],
averaged over $100$ replications. Errors are normalized so that the
Bayes estimator for each choice of $K$ and $\mu$ has error $1$.
Estimation methods are listed in the text. In the dense case, the
universal soft and hard thresholding methods are hidden because their
relative errors range from $4$ to $50$.}\label{fig:two-sided}\vspace*{-0.4pt}
\end{figure}

The mixture model was the best performer overall and in most of the
cases. Figures~\ref{fig:one-sided} and~\ref{fig:two-sided} show
the performance of the various methods relative to the Bayes estimator
for each scenario. The mixture model does a little better than the
other methods on sparse $\delta$ ($K=5$) and nearly achieves the
Bayes error for moderate and dense $\delta$ ($K=50,500$). Table
\ref{tab:ineff} gives the mean and median relative error over the
$24$ scenarios; the mixture model is often within $5\%$ of the Bayes
rule, and is the clear winner overall.

\begin{table}[b]
\tablewidth=290pt
\caption{Mean and median relative error for the
methods over
the simulation scenarios. The relative error is the average of the
squared error $\sum(\delta_{i}-\hat{\delta}_{i})^{2}$ over the $100$
replications, divided by the average squared error for the Bayes estimator}
\label{tab:ineff}
\begin{tabular*}{290pt}{@{\extracolsep{\fill}}lcc@{}}
\hline
\textbf{Method} & \textbf{Mean} & \textbf{Median}\\
\hline
\textbf{Mixture Model ($J=10$, $P=50$)} & \textbf{1.10} &
\textbf{1.04}\\[6pt]
Spline & 2.08 & 1.43\\
EBayesThresh & 1.70 & 1.39\\
FDR & 1.92 & 1.70\\
SUREShrink & 2.11 & 1.64\\
Universal hard & 3.60 & 2.47\\
Universal soft & 8.24 & 4.52\\
\hline
\end{tabular*}
\end{table}

\begin{figure}

\includegraphics{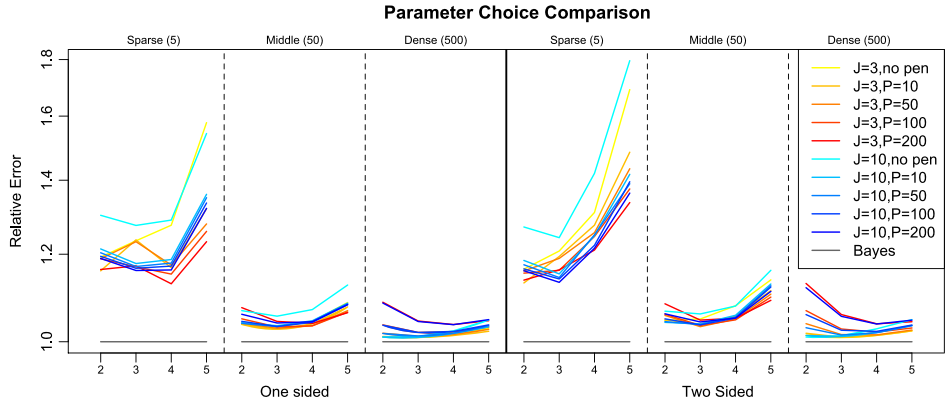}

\caption{Relative errors for various parameter
choices. Each
panel corresponds to one value of $K$ ($5$, $50$ or $500$). Within
each panel, $\mu$ increases from $2$ to $5$. The $y$-axis plots
the squared error [$\sum(\delta_{i}-\hat{\delta}_{i})^{2}$], averaged
over $100$ replications. Errors are normalized so that the Bayes
estimator for each choice of $K$ and $\mu$ has error $1$. The parameter
$J$ gives the number of groups in the mixture model, and $P$ is
a penalization parameter.}
\label{fig:param}
\end{figure}

The mixture model's performance is not because it is
fitting the true model---taking $J$ as low as $3$ gives the same
excellent performance (see Figure~\ref{fig:param}) even though the
data are certainly not generated from a three group normal mixture.
Neither is its performance due to careful tuning. Performance was
insensitive to parameter choice, as Figure~\ref{fig:param} shows.
The number of groups $J$ does not matter much and as long as there
is some penalization, the exact value of $P$ is not too important,
especially in the moderate and dense cases.

\subsection*{$\mathit{fdr}$ estimation}

We can also investigate the mixture model's $\mathit{fdr}$ and $\mathit{FDR}$ estimation
performance by examining a specific simulation. We generate $z_{i}\sim
\mathcaligr{N}(\delta_{i},1)$,
$i=1,\ldots,N=1000$. $950$ of the $\delta_{i}$ were $0$. The other
$50$ were drawn (once and for all) from a $\operatorname{Unif}(2,4)$ distribution.
Various methods were used to estimate the $\mathit{fdr}(z)=\P(\delta_{i}\ \mathit{null}|z_{i}=z)$
and $\mathit{FDR}(z)=P(\delta_{i}\ \mathit{null}||z_{i}|\geq z)$ curves based on $z_{i}$,
using either theoretical or empirical nulls:
\begin{itemize}
\item The normal mixture model with $J=3$ and $P=50$. For this simulation,
nearly null components were counted as null.
\item Locfdr, from \cite{Efron2008}. This fits the overall density using
spline estimation. It fits the empirical null by truncated maximum
likelihood (``ML'') or fitting a quadratic to $\log f$ near the
center (``CM'' for central matching). The implementation in the
R package ``locfdr'' was used.
\item Fdrtool, from \cite{Strimmer2008}. This fits the overall density
using the Grenander density estimator, and the empirical null by truncated
maximum likelihood. The implementation in the R package ``Fdrtool''
was used.
\end{itemize}
The whole simulation was run $100$ times, and the same random noise
was used for each method. The results are similar for other scenarios
and parameter choices; the simulation code is available in the Supplementary
Information online, and its parameters can be changed easily.

\begin{figure}

\includegraphics{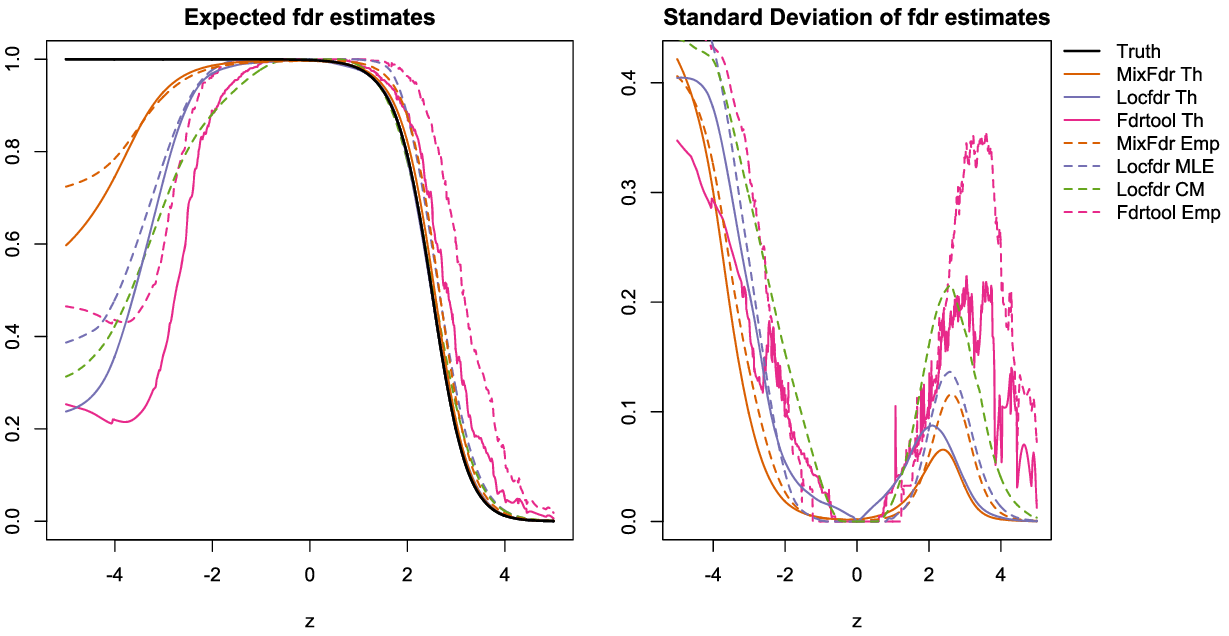}

\caption{$\E(\hat{\mathit{fdr}}(z))$ and $\operatorname{Sd}(\hat{\mathit{fdr}}(z))$
for various values of $z$ and the methods under consideration. ``Th''
means the theoretical null was used, while ``Emp'' means an empirical
null was used. Locfdr MLE and CM use the truncated maximum likelihood
and central matching empirical null estimates, respectively.}
\label{fig:fdrbias}
\end{figure}

\begin{figure}

\includegraphics{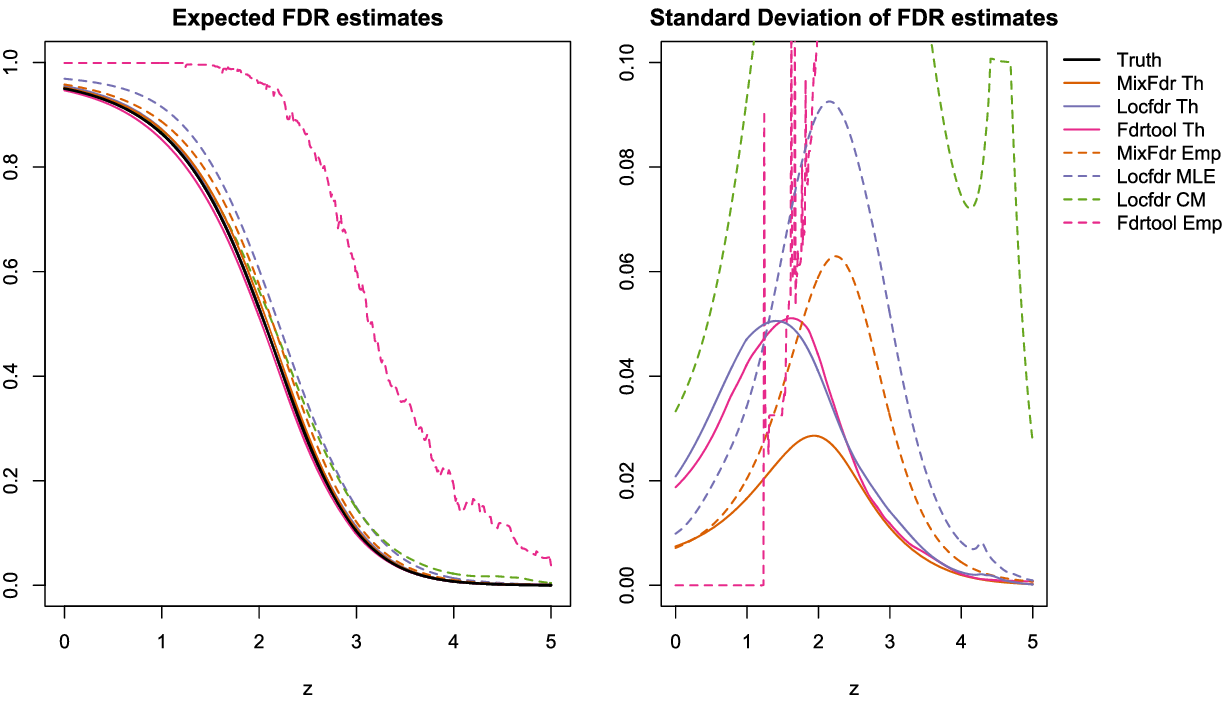}

\caption{$\E(\hat{\mathit{FDR}}(z))$ and $\operatorname{Sd}(\hat{\mathit{FDR}}(z))$
for various values of $z$ and the methods under consideration. ``Th''
means the theoretical null was used, while ``Emp'' means an empirical
null was used. Locfdr MLE and CM use the truncated maximum likelihood
and central matching empirical null estimates, respectively.}
\label{fig:FDRbias}
\end{figure}

The mixture model is probably the best $\mathit{fdr}$ and $\mathit{FDR}$ estimator,
but not by much, and the situation is more complicated than the effect
size situation. Figure~\ref{fig:fdrbias} shows the expectation and
standard deviation of $\widehat{\mathit{fdr}}(z)$ for the various methods. Fdrtool's
high bias and variance, and central matching's high variance, make
them poor $\mathit{fdr}$ estimators. This leaves Locfdr (and its ML empirical
null method) as the mixture model's only real competitor. Both methods
are nearly unbiased for positive $z$, and their bias for negative
$z$ is unlikely to be misleading. The mixture model is slightly more
stable than Locfdr, especially in the tails. Results for $\mathit{FDR}$ estimation,
seen in Figure~\ref{fig:FDRbias}, were similar.

%
%

\begin{figure}[b]

\includegraphics{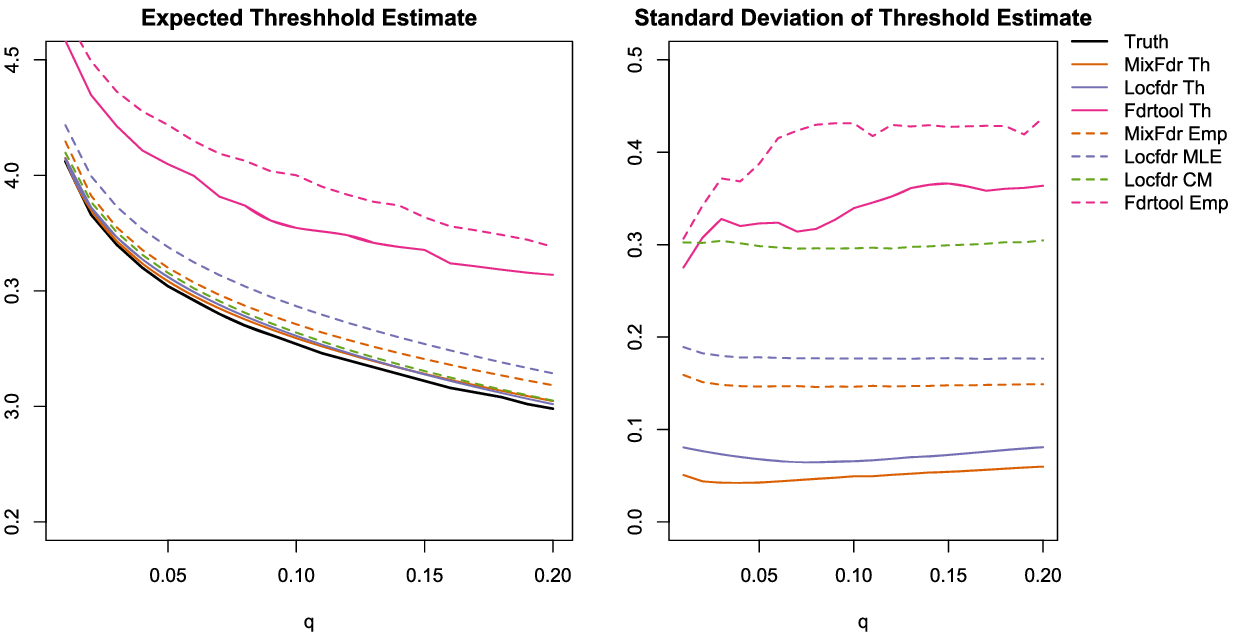}

\caption{Expectation and standard deviation of
rejection
threshold estimates $\hat{t}(q)$ for the various methods. The threshholds
are $\mathit{fdr}$ based.}
\label{fig:threshplot}
\end{figure}

\begin{figure}

\includegraphics{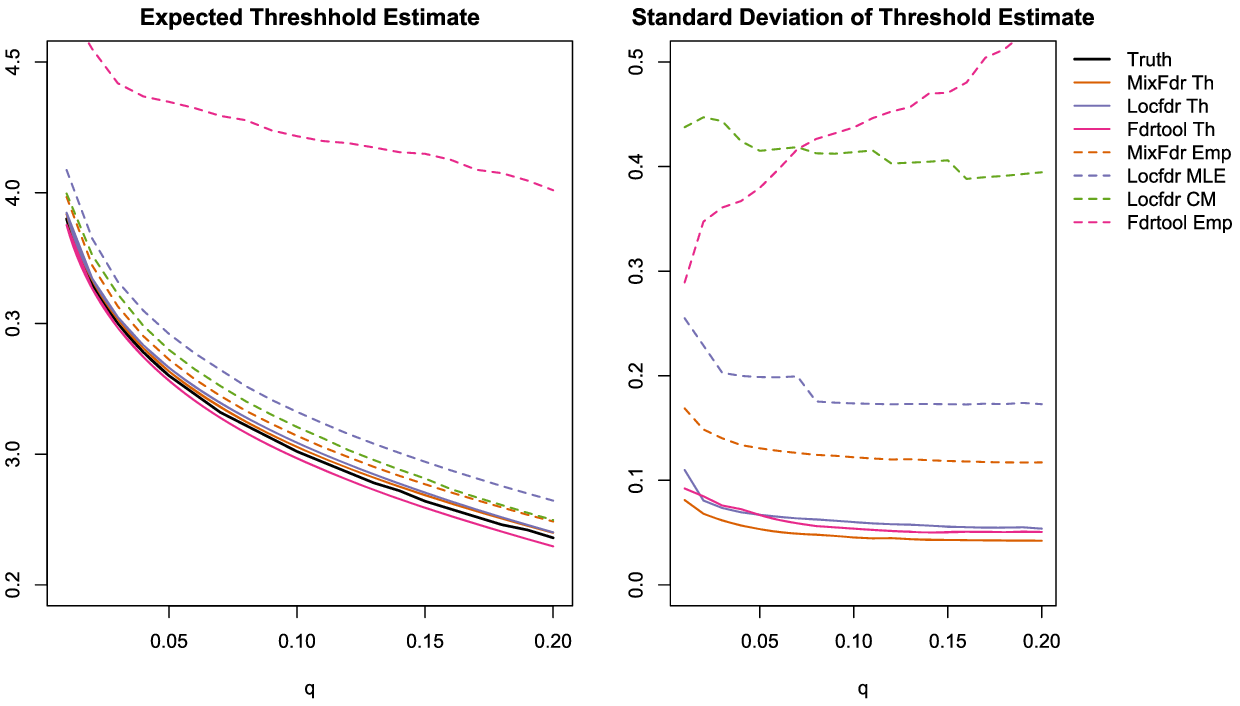}

\caption{Expectation and standard deviation
of rejection
threshold estimates $\hat{t}(q)$ for the various methods. The threshholds
are $\mathit{FDR}$ based.}
\label{fig:FDRthreshplot}
\end{figure}

The mixture model is nevertheless a little better, especially if we
need an empirical null. This is because of the way $\mathit{fdr}$ and $\mathit{FDR}$
estimates are usually used---we typically estimate $\mathit{fdr}(z)$, and
use our estimate to find rejection regions $\{z|\mathit{fdr}(z)\leq q\}$.
For moderate $q$ ($0.01$ to $0.2$), the rejection regions are in
the tails, where the mixture model is stabler. This means that the
mixture model is a stabler estimator of the rejection region than
Locfdr. In our simulation, the rejection region for a given $q$ corresponds
to rejecting all $z$ greater than some threshold $t(q)$. We can
use the $\mathit{fdr}$ estimation methods to estimate the rejection thresholds.
Figure~\ref{fig:threshplot} shows the expectation and standard deviation
of $\hat{t}(q)$ for the various methods. Both the mixture model and
Locfdr are nearly unbiased for the true threshold, for both theoretical
and empirical nulls. Locfdr, however, gives more variable threshold
estimates, especially with an empirical null. This makes the mixture
model a better choice for threshold estimation. This result held for
almost all parameter choices, and is true for $\mathit{FDR}$-based threshholds
as well (Figure~\ref{fig:FDRthreshplot}).

\section{Summary and extensions}\label{sec:Summary-and-Extensions}

To summarize, the mixture model approach is a simple, flexible and
accurate way to estimate $\mathit{fdr}$'s, $\mathit{FDR}$'s and effect sizes. It estimates
them together, instead of separately, and can fit an empirical null
if required. The method yields simple, interpretable models that can
be strongly parametric or quite nonparametric. The method has two
tuning parameters---the number of mixture components and the penalization.
It is quite insensitive to the first, and, for most purposes, the
second. We can choose the penalization by bootstrap calibration. Finally,
the method works for exponential families, and can easily accommodate
nuisance parameters. It is worth considering a few extensions of the
mixture model approach before we close.

The mixture model can be useful even when we are only interested in
$\mathit{fdr}$ or $\mathit{FDR}$ estimates. In these situations, the Brown--Stein model
imposes unnecessary restrictions on the marginal distribution of the
data; it makes sense to drop the model and work with the marginal
distribution directly, as much of the $\mathit{fdr}$ literature does
[\cite{Storey2002}, \cite{Efron2008}].
The mixture model approach can still be useful in these situations
- model the marginal as mixture and penalize the mixture proportions.
For example, for normal data, this amounts to modeling the marginal
as a normal mixture. This approach can incorporate empirical nulls
just as before. The mixture model's good performance should extend
to these approaches.

The mixture model can also be useful beyond exponential families.
Section~\ref{sec:The-Model} used exponential families for a convenient
definition of effect size, for their conjugate priors and for Lemma
\ref{lem:fdr}. None of these is central, so if we have data with
a natural notion of effect size, we can follow the mixture model's
approach: model the data using a prior on effect sizes, fit a mixture
prior by marginal maximum likelihood, then use the Bayes estimates
with the estimated prior. The loss of Lemma~\ref{lem:fdr} means that
there may be some identifiability issues, but the approach will often
still be successful.

\section*{Acknowledgments}

The author thanks Professor Robert Tibshirani and especially Professor
Bradley Efron for many discussions and useful comments. He also thanks
Professor Iain Johnstone for pointing out \cite{Brown2008} and providing
simulation code.

\begin{supplement}[id=suppA]
\exhyphenpenalty-10000
\sname{Supplement A}
\stitle{Model and Simulation Code\\}
\slink[doi]{10.1214/09-AOAS276SUPPA}
\slink[url]{http://lib.stat.cmu.edu/aoas/276/rcode.zip}
\sdatatype{.zip}
\sdescription{This file contains the batting average data, R code to
fit binomial normal
mixture models, and scripts to carry out the simulations and data
analysis performed in the paper.
The R package ``mixfdr,'' available from CRAN and the author's website,
has the code for the normal
mixture model.}
\end{supplement}

\begin{supplement}[id=suppB]
\exhyphenpenalty-10000
\sname{Supplement B}
\stitle{Fitting Details and Derivations\\}
\slink[doi]{10.1214/09-AOAS276SUPPB}
\slink[url]{http://lib.stat.cmu.edu/aoas/276/supplement.pdf}
\sdatatype{.pdf}
\sdescription{This document has more details on the EM algorithm used
to fit
the model and derivations of some posterior distribution formulas.}
\end{supplement}

\printaddresses


\begin{thebibliography}{99}

\bibitem[\protect\citeauthoryear{Abramovich et~al.}{2006}]{Abramovich2006}
 \textsc{Abramovich, F., Benjamini, Y., Donoho, D.~L.} and  \textsc{Johnston, I.~M.}
 (2006).
  Adapting to unknown sparsity by controlling the false
discovery rate.
 \textit{Ann. Statist.} \textbf{34}
584--653.
\MR{2281879}


\bibitem[\protect\citeauthoryear{Abramovich, Grinshtein and Pensky}{2007}]{Abramovich2007}
\textsc{Abramovich, F., Grinshtein, V.} and \textsc{Pensky, M.} (2007).
  On optimality of Bayesian testimation in the normal means problem.
 \textit{Ann. Statist.} \textbf{35}
2261--2286.
\MR{2363971}

\bibitem[\protect\citeauthoryear{Allison et~al.}{2002}]{Allison2002}
\textsc{Allison, D.~B., Gadbury, G.~L., Heo, M., Fernandez, J.~R.,
Lee, C.-K., Prolla, T.~A.} and \textsc{Weindruch, R.} (2002).
  A mixture model approach for the analysis of microarray gene
expression data.
 \textit{Comput. Statist. Data Anal.} \textbf{1} 1--20.
\MR{1895555}

\bibitem[\protect\citeauthoryear{Brown}{1971}]{Brown1971}
 \textsc{Brown, L.~D.} (1971).
  Admissible estimators, recurrent diffusions, and insoluble boundary
value problems.
 \textit{Ann. Math. Statist.} \textbf{42}
855--903.
\MR{0286209}

\bibitem[\protect\citeauthoryear{Brown}{2008}]{Brown2008}
\textsc{Brown, L.~D.}  (2008).
  In-season prediction of batting averages: A field test of empirical
Bayes and Bayes methodologies.
 \textit{Ann. Appl. Statist.} \textbf{2}
113--152.
\MR{2415597}

\bibitem[\protect\citeauthoryear{Cai, Jin and Low}{2007}]{Cai2007}
\textsc{Cai, T., Jin, J.} and \textsc{Low, M.} (2007).
  Estimation and confidence sets for sparse normal mixtures.
 \textit{Ann. Statist.} \textbf{35} 2421--2449.
\MR{2382653}

\bibitem[\protect\citeauthoryear{Donoho and Johnstone}{1994}]{Donoho1994}
\textsc{Donoho, D.~L.} and \textsc{Johnstone, I.~M.}  (1994).
  Ideal spatial adaptation by wavelet shrinkage.
 \textit{Biometrika} \textbf{81} 425--455.
\MR{1311089}

\bibitem[\protect\citeauthoryear{Donoho and Johnstone}{1995}]{Donoho1995}
\textsc{Donoho, D.~L.} and \textsc{Johnstone, I.~M.} (1995).
  Adapting to unknown smoothness via wavelet shrinkage.
 \textit{J. Amer. Statist. Assoc.}
\textbf{90} 1200--1224.
\MR{1379464}

\bibitem[\protect\citeauthoryear{Efron}{2004}]{Efron2004a}
\textsc{Efron, B.} (2004).
  Large-scale simultaneous hypothesis testing.
 \textit{J. Amer. Statist. Assoc.}
\textbf{99}
96--104.
\MR{2054289}

\bibitem[\protect\citeauthoryear{Efron}{2008a}]{Efron2008a}
\textsc{Efron, B.} (2008a).
  Empirical Bayes estimates for large-scale prediction problems.

\bibitem[\protect\citeauthoryear{Efron}{2008b}]{Efron2008}
\textsc{Efron, B.} (2008b).
  Microarrays, empirical Bayes and the two-groups model.
 \textit{Statist. Sci.} \textbf{23} 1--22.
\MR{2431866}

\bibitem[\protect\citeauthoryear{Efron}{2009}]{Efron2009}
\textsc{Efron, B.} (2009).
  Correlated $z$-values and the accuracy of large-scale statistical
estimates.


\bibitem[\protect\citeauthoryear{Efron et~al.}{2001}]{Efron2001}
\textsc{Efron, B., Tibshirani, R.,  Storey, J.~D.} and \textsc{Tusher, V.} (2001).
  Empirical Bayes analysis of a microarray experiment.
 \textit{J. Amer. Statist. Assoc.}
\textbf{96} 1151--1160.
\MR{1946571}

\bibitem[\protect\citeauthoryear{Jin and Cai}{2007}]{Jin2007}
\textsc{Jin, J}. and \textsc{Cai, T.} (2007).
  Estimating the null and the proportion of non-null effects in
large-scale multiple comparisons.
 \textit{J. Amer. Statist. Assoc.}
\textbf{102}
495--506.
\MR{2325113}

\bibitem[\protect\citeauthoryear{Johnstone and Silverman}{2004}]{Johnstone2004}
\textsc{Johnstone, I.~M.} and \textsc{Silverman, B.~W.} (2004).
  Needles and straw in haystacks: Empirical Bayes estimates of possibly
sparse sequences.
 \textit{Ann. Statist.} \textbf{4} 1594--1649.
\MR{2089135}

\bibitem[\protect\citeauthoryear{McLachlan and Peel}{2000}]{McLachlan2000}
\textsc{McLachlan, G.} and \textsc{Peel, D.} (2000).
 \textit{Finite Mixture Models}.
  Wiley-Interscience, New York.
\MR{1789474}

\bibitem[\protect\citeauthoryear{Meinshausen and Rice}{2006}]{Meinshausen2006}
\textsc{Meinshausen, N.} and \textsc{Rice, J.} (2006).
  Estimating the proportion of false null hypotheses among a large
number of independently tested hypotheses.
 \textit{Ann. Statist.} \textbf{34}
373--393.
\MR{2275246}

\bibitem[\protect\citeauthoryear{Muralidharan}{2009}]{Muralidharan2009}
 \textsc{Muralidharan, O.} (2009).
  Supplement to ``An empirical Bayes mixture method for false discovery
rate and effect size estimation''.
 \textit{Ann. Appl. Statist.}
DOI: \href{http://dx.doi.org/10.1214/09-AOAS276SUPPA}{10.1214/09-AOAS276SUPPA},
DOI: \href{http://dx.doi.org/10.1214/09-AOAS276SUPPB}{10.1214/09-AOAS276SUPPB}.

\bibitem[\protect\citeauthoryear{Newton et~al.}{2004}]{Newton2004}
\textsc{Newton}, M.~A., \textsc{Noueiry},  A., \textsc{Sarkar},  D. and \textsc{Ahlquis}, P. (2004).
  Detecting differential gene expression with a semiparametric
hierarchical mixture method.
 \textit{Biostatistics} \textbf{5} 155--176.

\bibitem[\protect\citeauthoryear{Pan, Lin and Le}{2003}]{Pan2003}
\textsc{Pan, W., Lin,  J.} and \textsc{Le, C.~T.} (2003).
  A mixture model approach to detecting differentially
expressed genes
with microarray data.
 \textit{Functional and Integrative Genomics} \textbf{3}
117--124.

\bibitem[\protect\citeauthoryear{Pensky}{2006}]{Pensky2006}
\textsc{Pensky, M.} (2006).
  Frequentist optimality of Bayesian wavelet shrinkage rules for
Gaussian and non-Gaussian noise.
 \textit{Ann. Statist.} \textbf{34}
769--807.
\MR{2283392}

\bibitem[\protect\citeauthoryear{Robbins}{1954}]{Robbins1954}
\textsc{Robbins, H.} (1954).
  An empirical Bayes approach to statistics.
  In  \textit{Proc. Thrid Berkeley Sympos. Math. Statist. Probab.} \textbf{1} (J. Neyman, ed.)  157--163.
Univ. California Press, Berkeley, CA.
\MR{0084919}

\bibitem[\protect\citeauthoryear{Storey}{2002}]{Storey2002}
\textsc{Storey, J.~D.} (2002).
  A direct approach to false discovery rates.
 \textit{J. Roy. Statist. Soc. Ser. B}
\textbf{64} 479--498.
\MR{1924302}

\bibitem[\protect\citeauthoryear{Strimmer}{2008}]{Strimmer2008}
\textsc{Strimmer, K.} (2008).
  A unified approach to false discovery rate estimation.
 \textit{BMC Bioinformatics} \textbf{9} 303.

\end{thebibliography}
\end{document}